\documentclass[11pt]{article}

\usepackage[english]{babel}

\usepackage[letterpaper,margin=2.5cm]{geometry}

\usepackage{setspace}
\usepackage{amsmath}
\usepackage{graphicx}
\usepackage[hidelinks]{hyperref}
\usepackage{booktabs}
\usepackage{authblk}
\usepackage[symbol]{footmisc}
\usepackage[super]{cite}
\usepackage{xurl}

\usepackage{caption}
\usepackage{subcaption}
\date{}

\title{A field experiment of social influence and behavioral contagion \\ with bots on Reddit}

\author[a]{Hiroki Oda}
\author[a,b]{Kinga Makovi}
\author[c,d]{Taha Yasseri}
\author[a,e,*]{Milena Tsvetkova}

\affil[a]{London School of Economics and Political Science, London, UK}
\affil[b]{New York University Abu Dhabi, Abu Dhabi, UAE}
\affil[c]{Centre for Sociology of Humans and Machines, Trinity College Dublin, Dublin, Ireland}
\affil[d]{School of Mathematics and Statistics, University College Dublin, Dublin, Ireland}
\affil[e]{Complexity Science Hub, Vienna, Austria \vspace{0.5cm}}
\affil[*]{To whom correspondence should be addressed: m.tsvetkova@lse.ac.uk}

\begin{document}
\maketitle

\begin{abstract}
\noindent Recent advances in AI have heightened scholars’ and policy makers’ concern with social influence and behavioral contagion in online communities. We conduct a field experiment on Reddit to investigate the extent to which online users are susceptible to positive behavioral stimuli from other users and artificial agents. We let apparent human and bot accounts give symbolic awards to users with one of four rationales: praising the recipient’s logical argument, emotional sensitivity, or moral integrity, or explaining that the award resulted from a random draw in a lottery. We evaluate how the different rationales for the award affect the recipients’ subsequent behavior on the platform in terms of volume, impact, and content, as well as the further behavioral contagion to other users. We find that awards do not increase user activity and downstream impact, and awards from bots with the lottery rationale can in fact reduce them. Nevertheless, awards encourage direct communication between users. These findings highlight the possible resilience of online users to simple behavioral manipulation from platform algorithms and artificial agents, but not necessarily to more sophisticated schemes that simulate human conversation. Transparently labeling automated agents remains essential for ethical and effective platform governance.
\end{abstract}

\section{Introduction}

From fake accounts on social media and generative-AI chatbots to self-driving vehicles on the roads and trading algorithms in global markets, modern society has become a hybrid social system of humans and intelligent machines who interact with and influence each other \cite{pedreschi2023, rahwan2019, tsvetkova_new_2024}. Under certain conditions, intelligent machines may affect how humans make decisions and act, and even if weak, this influence can cascade in social networks, producing palpable macro-level consequences. While some research finds that humans treat and respond to machines similarly to other humans \cite{mutzner_normative_2026, nass_computers_1994, nass_machines_2000}, other work identifies notable differences between human-human and human-machine interactions \cite{chugunova2022, makovi_rewards_2025, march_strategic_2021}. To grasp today’s social world, we need to understand how intelligent machines influence and affect human behavior and what consequences this influence has for social groups, networks, communities, and society at large.

In this study, we investigate whether social influence and contagion differ between human and artificial influencers for different rationales of action through a field experiment in a large online community with an endemic bot population. We create apparent bot and human user accounts on the online news sharing and discussion forum Reddit and let them give symbolic awards to users with different stated justification, presupposing evaluation of rationality and logical reasoning, adherence to a procedure, albeit arbitrary, understanding of emotional states, or capacity for moral judgement. We evaluate how the randomized interventions affect the volume, impact, and content of subsequent contributions by the award recipients, as well as any spill-over effects to other users.

There are several arguments why we expect influence from machines versus humans to differ for different action framings. Prior research suggests that humans perceive and judge intelligent machines differently from how they perceive and judge other humans. To begin with, humans do not ascribe moral agency to machines \cite{gray2007, krach2008, mccabe2001, zhang2022} and perceive them to act with lower intentional capacity and with less bias compared to humans. As a result, while humans judge other humans by their intentions, they judge machines by their performance; thus, they would blame other humans more for (intentional) lack of fairness but blame machines more for (accidental) harm \cite{hidalgo2021}. For instance, experimental participants are more likely to pay to punish free riding by perceived bots than by other humans \cite{chugunova2022, makovi_rewards_2025, march_strategic_2021}. Further, humans trust machine rationality and precision more. Thus, when facing potential losses, humans are more likely to delegate decisions to AI than to other humans \cite{candrian2022}. However, if a machine commits a mistake, especially in an objective area or an analytical task, humans are less willing to forgive and quickly lose their trust \cite{dietvorst2015}. Finally, humans exhibit a narrower emotional spectrum towards machines than towards other humans, reacting less positively but also less negatively \cite{adam2018, chugunova2022}. For instance, humans react with lower and flatter levels of social emotions such as gratitude, anger, pride, and guilt to bots’ investment decisions in a trust game compared to humans’ decisions \cite{schniter2020}.

Humans may perceive machines differently, but they are still susceptible to their influence when making decisions or solving problems \cite{kobis2021}. According to the CASA (computers as social actors) paradigm in psychology, in many situations, humans respond to machines similarly to how they do to other humans. For instance, people reciprocate kind acts by computers \cite{fogg_how_1997}, treat them as politely as they treat humans \cite{nass_computers_1994}, consider them as competent, and intend to follow them as much \cite{edwards_is_2014}, but also apply gender and racial stereotypes to them \cite{bazazi_ais_2025, cui_gender_2026, nass_machines_2000, siegel_persuasive_2009}. Robots can cause informational and normative conformity in people \cite{salomons_humans_2018, salomons_minority_2021}, AI advice can corrupt humans’ moral judgment \cite{krugel_chatgpts_2023, leib2021}, while social media bots can make users imitate the sentiment and words of their posts \cite{ma2020}. Humans tend to trust algorithmic advice more than advice coming from a human or a human crowd, especially for more difficult tasks \cite{bogert_humans_2021, logg_algorithm_2019}. Yet, humans may be reluctant to adopt algorithmic advice if they perceive a threat to their decision control or have difficulty understanding and internalizing the advice \cite{burton_systematic_2020, mahmud_what_2022}. 

Even if machine-human influence is weaker, it can still cascade through human-human interactions in social networks. Influence can spread directly between interactants via “pay-it-forward” chains or indirectly from interactants to observers \cite{tsvetkova_social_2014}. Recommendation and ranking algorithms can also indirectly trigger contagion by promoting and reinforcing behaviors and information \cite{pescetelli2022}. In this way, small individual-level effects can produce chain reactions that spiral into large group-level consequences \cite{fowler2010, leskovec2007, watts_simple_2002}.

In fact, in certain situations, weak individual-level effects may offer an advantage. For instance, covert social media bots exert only weak, slow, and unobtrusive influence on humans but this prevents them from alienating their followers and their followers’ friends compared to more pushy and assertive users; as a result, their messages can spread farther and sway public opinion and exacerbate political polarization \cite{keijzer2021, ross2019}. Similarly, even though robots and bots evoke flatter emotional reactions from individual humans, they can trigger emotional contagion in groups, increasing social interactions, laughter, sharing, and mutual support \cite{strohkorbsebo2018, traeger2020}. Overall, persistence, strategic placement, and sheer numbers can compensate for bots’ weak direct influence on humans to produce significant collective influence \cite{pescetelli2022, shao2018, stella2019, vosoughi2018}.

In sum, machines exert social influence on humans that can spread and grow. However, since humans perceive and judge machines differently from other humans, the strength of this influence and its overall impact may depend on the behavior and its apparent rationale. We thus distinguish between influence for decisions made by bots (B) versus humans (H) using one of four different rationales: rational/logical (LOG), random/arbitrary (RAN), emotional/sentiment-driven (EMO), and moral/normative (MOR). Although prior findings are mixed, we guide our research with directed hypotheses. First, we assume that social influence is real and regardless of their source or rationale, others’ decisions make a difference compared to no decisions (control C). We expect that bots have a stronger influence when they employ a logic-based or a procedurally fair rationale such as random selection than emotion-based or moral-based rationales, which are outside of their perceived competencies:  B\textsubscript{LOG, RAN} $>$ B\textsubscript{EMO, MOR}; for humans, we expect the reverse: H\textsubscript{LOG, RAN} $<$ H\textsubscript{EMO, MOR}. We extend this reasoning to predict that bots will be more influential than humans for logic-based or random-selection rationales (B\textsubscript{LOG, RAN} $>$ H\textsubscript{LOG, RAN}) but less influential for emotion-based or moral-based rationales (B\textsubscript{EMO, MOR} $<$ H\textsubscript{EMO, MOR}). It remains an open question whether human emotions and morals trump machine rationality and procedural fairness: B\textsubscript{LOG, RAN} $<$ H\textsubscript{EMO, MOR}.

\section{Materials and Methods}

We conducted a field experiment on Reddit, a popular news aggregation, content rating, and discussion website founded in 2005. Bots are endemic to the platform and communicate rules, moderate contributions, augment functionality (e.g., for mobile users), or post content, ranging from comic and playful posts by evident automated accounts such as haiku\_robot and ObamaRobot, to trolling and provocative comments by undercover social bots \cite{massanari2016}. The Reddit community encourages developers to follow the crowdsourced protocol for deploying Reddit bots known as the ``Bottiquette'' and specifically, to include the word ``bot'' in the bot account username \cite{hurtado2019}. Reddit largely relies on peer monitoring and sanctioning within smaller sub-communities (subreddits), which have fostered a more open and positive relationship between users and bots, similar to the productive collaboration observed on Wikipedia \cite{geiger2011, halfaker2012} and in contrast to the distrust and manipulation plaguing social media sites like X and Facebook \cite{ferrara_rise_2016, gorwa_unpacking_2020}.

We create a typical human account and a clear bot account on Reddit and use them to give a non-anonymous award to a post or comment with a short message to the user stating a specific rationale. An award appears as a badge below the awarded contribution and is visible to any other user. For non-anonymous awards, the recipient also receives a notification with the awarding account's username and message. To ensure interventions are credible and more impactful, we pre-select non-controversial posts and comments of a certain minimum length that are contributed by relatively inexperienced users, and we treat them early, before others have voted and commented on them.

We obtained ethical approval for the experimental design (LSE Ref.490807) and pre-registered the hypotheses.\footnote{Available at: \url{https://aspredicted.org/n5v5-jwg5.pdf}} The anonymized and aggregated experimental data and the analysis scripts will be made publicly available upon publication.

\subsection{Sample selection}

We first identify a set of popular and active general-interest subreddits on non-sensitive topics with primarily text-based posts and/or comments. We begin with the top 250 subreddits by membership size in the “Best of Reddit” ranking and exclude subreddits with posts that involve images, gifs, or videos, subreddits on special-interest topics such as mathematics, online games, and technology brands, and subreddits dedicated to sensitive topics such as politics, mental health, and self-help. We end up with 12 subreddits about personal and general advice, and knowledge and information (Table~\ref{sec:table_s1}).

From this set of subreddits, we continuously query for promising recent contributions that do not concern sensitive topics. Specifically, we screen for contributions posted within 12 hours of each screening run with 1–5 comments and 2–6 upvotes for posts and 2–21 upvotes for comments (comments tend to get more attention in general), no downvotes, and a reasonable length of 300–3000 characters. We then exclude potentially problematic contributions on sensitive topics, first with a dictionary-based filter and then, with a GPT-4 screen (see ~\ref{sec:gpt4_screen} Sample Selection with GPT-4 in SI). GPT-4 is capable of uncovering implicit topics, which dictionary methods miss, and has been proven to achieve fair levels of agreement with humans on text classification tasks \cite{ziems_can_2024}. 

Finally, from the resulting set of contributions, we select those by relatively inexperienced users without moderation privileges who created their account at least 30 days prior to the time of treatment and have 10–499 comments and 0–99 posts contributed so far. Based on a pilot test with four treatment conditions of 40 users each, we estimated an average effect size of 0.5, a standard deviation of 0.1, and a difference of 0.03 between treatments. To detect it at the 0.02 significance level (due to multiple tests) with a power of 85\%, we require about 250 observations per condition. After removing several problem cases that slipped through our filters, we end up with 2,442 observations in total with 263–275 per condition.

\subsection{Experimental procedure}

After selecting a contribution, we randomly assign it to one of the nine treatment conditions: four rationales sent from two user accounts, alongside a no-intervention control condition (contributions that were selected based on the same criteria and were simply monitored). We use the username, name, avatar, and background image to present one of the user accounts as a typical human user and the other as an apparent bot. In either case, we employ generic text and images, without hints to gender, age, or any other demographic information. We also complete the profile description with a short message that announces the user’s intention to give awards as a way to contribute to the community (see Fig.\ref{fig:figs1} in the Supplementary Information). 

For the eight treatment conditions that require an intervention, we use the respective research account to give a non-anonymous Reddit “Diamonds are Forever” award to the poster (worth 50 Gold coins or \$1.00) with a message to the recipient. Although awards cost money to send on Reddit, recipients cannot monetize them directly unless they have signed up for the Contributor Program, which is unlikely to be the case for the relatively inexperienced users we target. The awards we thus give are largely symbolic and reputation-boosting. The award is visible to other users but the award giver’s identity and message are only visible to the award recipient. Restricted to 100 characters by the platform, the message we send to the recipient has the format “[Identity hint]. [Rationale]. Keep it up!”, where the identity hint is “Hi, Reddit stranger” for (H) and “I’m a bot, beep boop” for (B), and the rationale is “I like your clear reasoning and logic” for (LOG), “I give away awards and your post won today’s lottery” for (RAN), “I like your emotional tone and empathy” for (EMO), and “I like your moral intuition and integrity” for (MOR).

Nine days after treatment, we download the treated user’s complete activity history, which enables us to estimate before- and after-treatment activity levels, content, interaction patterns, and impact. To study spillover effects, we record the history of all the post-treatment contributions by the treated user; for each contribution, this gives us the complete thread content and hierarchy with timestamps and contributors’ usernames. We also download any direct private messages sent by the treated users to the research accounts. Data collection took place from April 2025 until January 2026 using the Reddit API.\footnote{Source: \url{https://developers.reddit.com/docs/capabilities/server/reddit-api}}

\subsection{Statistical analysis}

The analyses estimate and compare the direct behavioral, emotional, and linguistic influence on the recipient, as well as any indirect influence on other users. We test for direct effects with the number of private thank-you messages and the volume (number and total text length) and content (positive sentiment and the lexical prominence of the award rationale) of the treated user’s post-treatment contributions. Contributions include both original posts and comments, but given users typically comment far more frequently than they post on Reddit, comments constitute the large majority of contributions in our data. 

We investigate indirect effects, or impact, with the number of votes, the number of comments or replies, and the total text length of the comments/replies to the treated user’s post-treatment contributions. We estimate the number of votes with the formula $2ur/(2r-1)$ for posts, where $u$ is the upvote score and $r$ is the upvote ratio available from the Reddit API. Since the upvote ratio is not available for comments, we estimate their votes with $|u|$. We measure positive sentiment using the metric “Positive tone” from the dictionary-based LIWC-22 software \cite{tausczik_psychological_2010}.\footnote{Source: \url{https://www.liwc.app/}, Python library: \url{https://www.liwc.app/help/cli}}  We measure the lexical prominence of the LOG and EMO rationales with the Analytical Thinking summary measure and the Emotion metric from the same software, and we measure the lexical prominence of the MOR rationale with the frequency of moral foundations words according to a publicly available dictionary \cite{kraft_replication_2017}.\footnote{Source: \url{https://dataverse.harvard.edu/dataset.xhtml?persistentId=doi:10.7910/DVN/WJXCT8}, R library: \url{https://content-analysis-with-r.com/4-dictionaries.html}} The positive tone, emotion, and moral language scores stand for the percentage of words in the respective dictionary, while the Analytical Thinking measure is a standardized score produced by a proprietary algorithm and converted to percentiles based on the area under a normal curve. Since the measures are unreliable for short text, we report results for users who contributed at least 50 words each in the before-treatment and after-treatment observation periods. This left 545 observations, between 50 and 68 per treatment group.

Due to bursty activity and positive feedback effects on Reddit, the measures we use exhibit long-tailed distributions, with a mode of zero but also several extreme outliers. To account for high between-individual variability, we measure within-individual changes: we calculate the difference between the treated user’s metric ($M$) for contributions made after treatment and the equivalent metric estimated for contributions made in the week before treatment ($\Delta M = M_{\text{after}} - M_{\text{before}}$). Ordinal data is difficult to visualize, so for the plots, we log-rescale the measures before taking the difference between before and after treatment: we plot the difference of the logarithms $\log(M_{\text{after}}) - \log(M_{\text{before}})$, which is equivalent to the logarithm of the ratio, or $\log(M_{\text{after}} / M_{\text{before}})$. We expect the effects from receiving an award to be immediate and short-lived. Hence, we statistically test the differences in $\Delta M$ for contributions made 12 hours after treatment. In the SI, we also include visualizations of the differences measured over a five days window after treatment (Fig.~\ref{fig:figs2}--\ref{fig:figs5}). 

We use the Mann-Whitney U test to statistically test the differences between treatments. This non-parametric mean-comparison test is suitable for unknown distributions and ordinal data, hence appropriate for the non-normal distributions we are dealing with. As per the hypotheses, we compare the H and B treatments for each LOG, RAN, EMO, and MOR rationales and then compare the rationales for each of the H and B treatments. To establish the size of effects, we additionally include the no-intervention control group. We originally planned to correct for multiple comparisons with the Holm-Bonferroni method \cite{holm_simple_1979}; however, given the surprising null and backfire effects, we report uncorrected p-values here. We discuss the strength of our findings at length in the conclusion. 

\section{Results}

\subsection{Direct effects on the treated user}

\begin{figure}[ht]
\centering
\includegraphics[width=\linewidth]{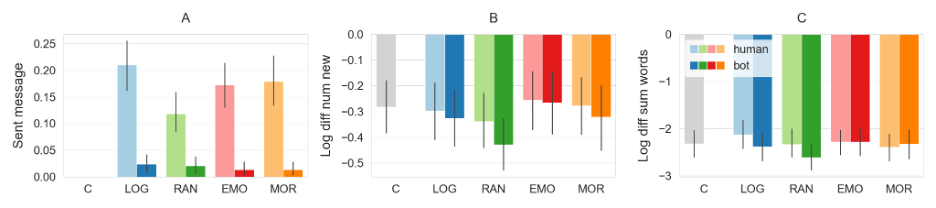}
\caption{\textbf{The direct effects of award account and rationale on the award recipient's A) private messages and B) number and C) combined text length of new contributions made in the 12 hours after receiving the award.} Users are more likely to reply to a human account than a bot account and when sent by a human account, the random-lottery rationale elicits fewer replies than the other rationales. Awards do not appear to increase activity and in fact, awards by a bot account with a random-lottery rationale have a negative effect on the number and combined text length of subsequent contributions. Panel A shows the fraction of award recipients who sent a private message to the awarding account. Panels B and C show mean estimates and 95\% CI for $\log(\text{after-treatment}) - \log(\text{before-treatment})$ of the measure; note, however, that statistical significance was established with non-parametric tests on $(\text{after-treatment} - \text{before-treatment})$.}
\label{fig:fig1}
\end{figure}

Many of the award recipients reply to the awarding account with a private message, typically expressing gratitude and friendliness. Users are significantly more likely to reply to a human than a bot (Fig.~\ref{fig:fig1}A). The responses to bots are too infrequent to compare between rationales. However, using a two-proportion z-test, we find significantly fewer replies to humans for the random rationale than the logic ($z = 3.01$, $p = 0.003$) and the moral ($z = 2.08$, $p = 0.038$) rationales, with the emotional rationale slightly below statistical significance at the usual level ($z = -1.86$, $p = 0.063$). This suggests that non-anonymous awards may help initialize direct contact among human users, especially when the award message contains praise. The observed differences also validate the salience of our manipulations – users noticed who sent them an award and why. 

Despite this, we find no evidence that awards increase the volume of activity: compared to the control group, none of the treatments have a statistically significant positive effect on the number of contributions and total words. In fact, the lottery awards sent by bots seem to backfire (Fig.~\ref{fig:fig1}B-C). Compared to the control group, recipients of awards sent by a bot account with a random rationale make fewer contributions ($U = 32802.5$, $p = 0.027$) with a shorter combined word length ($U = 32077.5$, $p = 0.009$): B\textsubscript{RAN} $<$ C. Users who receive awards from bots with the random rationale make significantly fewer contributions afterwards compared to users who receive awards from bots with the emotional rationale ($U = 32778.5$, $p = 0.031$) and from humans with the emotional ($U = 39388.0$, $p = 0.023$) and moral ($U = 40356.5$, $p = 0.010$) rationales: B\textsubscript{RAN} $<$ (B\textsubscript{EMO}, H\textsubscript{EMO, MOR}). Their contributions are also significantly shorter in total length when compared to users who receive awards from bots with the logical ($U = 41406.5$, $p = 0.016$) and emotional ($U = 32058.0$, $p = 0.011$) rationales and from humans with the logical ($U = 40957.5$, $p = 0.030$) and moral ($U = 39794.5$, $p = 0.025$) rationales: B\textsubscript{RAN} $<$ (B\textsubscript{LOG, EMO}, H\textsubscript{LOG, MOR}).

\begin{figure}[ht]
\centering
\includegraphics[width=\linewidth]{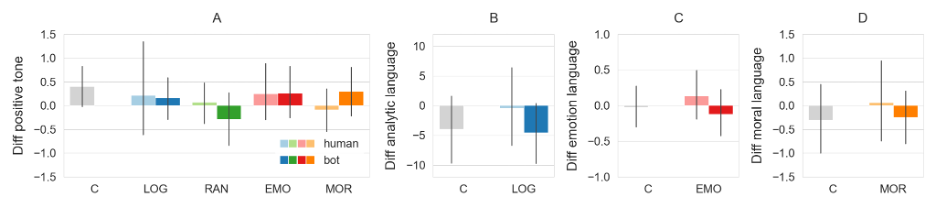}
\caption{\textbf{The direct effects of award account and rationale on the A) positive tone B) analytic language, C) emotion-related language, and D) moral language of the contributions the award recipient made in the 12 hours after receiving the award.} Awards do not appear to change the content in posts. The only statistically significant effect is that awards by a bot account with a random-lottery rationale decrease the positive tone of subsequent contributions compared to the control condition. Panels show mean estimates and 95\% CI for $(\text{after-treatment} - \text{before-treatment})$ of the measure.}
\label{fig:fig2}
\end{figure}

The expected effects on the content of the recipients’ contributions are also largely null, with some possible backfire effects (Fig.~\ref{fig:fig2}). The only statistically significant result is that recipients who receive an award by a bot with the random-lottery rationale decrease the positive tone of their posts compared to the control group in their post-treatment contributions: B\textsubscript{RAN} $<$ C ($U = 1193.5$, $p = 0.037$; Fig.~\ref{fig:fig2}A).

\subsection{Indirect effects on other users}

\begin{figure}[ht]
\centering
\includegraphics[width=\linewidth]{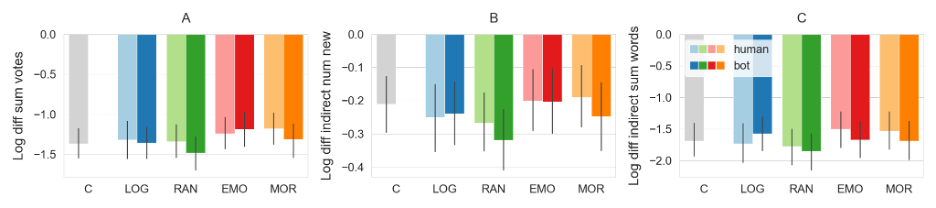}
\caption{\textbf{The indirect effects of award account and rationale on the A) total votes and the B) number and C) combined text length of the comments/replies to contributions the award recipient made in the 12 hours after receiving the award.} Awards increase the impact of the award recipient's subsequent contributions only when they come from human accounts with the moral rationale. The figure shows mean estimates and 95\% CI for $\log(\text{after-treatment}) - \log(\text{before-treatment})$; statistical significance was established with non-parametric tests on $(\text{after-treatment} - \text{before-treatment})$.}
\label{fig:fig3}
\end{figure}

For an award by a human account with the moral rationale, the recipient’s subsequent contributions garner significantly more votes compared to the control group ($U = 40029.5$, $p = 0.048$) and to awards by human accounts with the logic and random rationales ($U = 32648.5$, $p = 0.031$ and $U = 32648.0$, $p = 0.036$ respectively; Fig.~\ref{fig:fig3}A). To put it in notation, H\textsubscript{MOR} $>$ (C, H\textsubscript{LOG, RAN}). Part of these results are confirmed when we look at the text length of the comments and replies that the recipients’ contributions garner: longer for H\textsubscript{MOR} compared to users in the H\textsubscript{LOG} ($U = 31851.5$, $p = 0.009$) and H\textsubscript{RAN} ($U = 32768.5$, $p = 0.042$) conditions (Fig.~\ref{fig:fig3}C). Since the contributions are not more numerous and lengthier, nor with significantly more positive tone or moral language, the explanation for their success must be another aspect of quality that we do not measure. Apart from H\textsubscript{MOR}, however, there is no additional evidence that awards increase the impact of the award recipient’s contributions and that these effects differ by whether the award comes from a human or a bot.
 
\subsection{Effects on “activated” users}

Experimental manipulations online tend to be weak because users are distracted and overwhelmed by information, incessant notifications, vivid images, etc. We do not know whether users saw their award and read the award rationale, except for those who acknowledged it by sending back a private reply. Replicating the analyses for this subset offers higher internal validity but lowers the statistical power and restricts the conclusions to the human treatments only. The results suggest more clearly positive effects from awards (Fig.~\ref{fig:figs5}). The results provide additional evidence for the effect of awards with the moral rationale from human accounts: they increase the number of contributions ($U = 7939.0$, $p = 0.042$), which in turn obtain more votes ($U = 7964.0$, $p = 0.038$) compared to the control group.

\section{Discussion}

People’s susceptibility to social influence is a well-established phenomenon, exploited by political canvassers, volunteer projects, policy nudges, and interest-based communities, as well as marketeers and scammers. With respect to positive, beneficial effects, it has been shown that informal awards to online contributors increase effort and productivity, as well as further recognition by other users \cite{burtch_how_2022, restivo_experimental_2012, willer_groups_2009}. Here, we conducted a field experiment on Reddit to investigate how the effect of informal awards might be changing in the age of AI, when bots operate alongside humans. In contrast to previous findings, we found weak to no effects of awards on activity and impact, whether the award came from a human or a bot. In fact, we found a backfire effect from awards given at random by a bot. We detected a small positive effect on the impact of awards given by human accounts that praise the recipient’s moral reasoning. However, we note that this effect disappears if we enforce more stringent significance tests that control for multiple comparisons. 

Nevertheless, we found that non-anonymous awards with messages help initiate direct contact between users, and this effect is only meaningful for humans. Even then, the rationale matters: users are more likely to respond with a private message to a user with substantive praise than to a user who admits giving away awards at random.  

Our null findings on any positive effects from awards on effort contradict the results of a previous Reddit field experiment \cite{burtch_how_2022}. That study used anonymous awards before the platform-wide overhaul of the award system and a three-times larger sample for one control and one treatment condition. Hence, the differences in findings can be attributed to several factors. Anonymous awards are likely more common, so our non-anonymous awards with messages could have been interpreted as incentivized invitations for private conversations, more than as a recognition of contributions. In mid-2023, Reddit interrupted its old reward system, a decision that was widely unpopular, and it took a couple of years before the new, more heavily monetized version we used was rolled out.  Modifications have been ongoing. For instance, halfway through the data collection, the platform changed how users receive notifications for awards. It is thus possible that the changes have devalued the meaning and use of awards on Reddit, and this is why we observe null effects.    

It is also possible that our experiment is underpowered due to the nine tested conditions. Reddit presents particular challenges in this regard since activity tends to be bursty and reactive. Thus, high between-individual variability and high exogenous unpredictability might be washing out what is already expected to be a weak effect from a symbolic online intervention. Collecting more data would be ethically problematic because our interventions do not appear to yield benefits. We are also deeply aware that we conducted the experiment without the participants’ informed consent and with a mild form of deception (see Ethical Considerations in the Supplementary Information). Although Reddit has been generally welcoming to bots, the spread of large-language models and their documented abuse \cite{ogrady_unethical_2025} has shifted attitudes and the platform has started taking a more aggressive stance. Thus, for instance, six months after we started interventions, two of our “human” accounts got shadow banned, while the others continued without disruptions. 

Overall, our experiment shows that simple ways to manipulate humans may not always work, or possibly, have recently stopped working. On Reddit, symbolic non-anonymous awards appear to have neither direct nor spillover benefits on activity and productivity but seemingly serve to start private conversations, potentially kindling online friendships. Automated awards by bots can in fact backfire, reducing user engagement at least temporarily. There is a large variation across platforms in terms of their normative values and emergent culture, so our findings on Reddit may not generalize to other online communities. Yet, we could also speculate that the aggressive monetization of platforms and the spread of AI agents has made online users more cautious and cynical, while strengthening the meaning and value of direct connection and immediate communication. If generalized trust is eroding online, low-effort peer-to-peer interactions such as symbolic awards, upvotes, likes, re-shares may start losing effectiveness. Unfortunately, nefarious opinion-manipulation schemes will likely respond to this trend by simulating deeper human interactions in more elaborate ways. Since users react to what they believe to be other human users differently than to bots, transparent labels for artificial agents are imperative online. This will not only prevent malevolent exploitation but also strengthen user participation, engagement, and trust.

\section{Acknowledgments}

This project resulted from a grant funded by the LSE Research Impact and Support Fund 2024. The work was additionally supported by European Research Council grant HUMANET No.101170272. K.M.~acknowledges support from the NYUAD Center for Interacting Urban Networks (CITIES), funded by Tamkeen under the NYUAD Research Institute Award CG001. T.Y.~was funded by Research Ireland under grant number IRCLA/2022/3217, ANNETTE (Artificial Intelligence Enhanced Collective Intelligence). T.Y. also thanks Workday Inc for support.

\clearpage
\bibliographystyle{unsrt}
\bibliography{references}

\clearpage
\setcounter{section}{0}
\setcounter{figure}{0}
\setcounter{table}{0}
\renewcommand{\thesection}{S\arabic{section}}
\renewcommand{\thefigure}{S\arabic{figure}}
\renewcommand{\thetable}{S\arabic{table}}

\begin{center}
{\LARGE\textbf{Supplementary Information}}\\[0.5em]
\end{center}

\section{Data Recording and Tracking}

We select and treat posts and comments every 30 minutes. Upon intervention and then every two hours for the next 48 hours, we record the treated contribution’s performance (upvote count, upvote ratio, and comment count), the parent post’s performance in the case of treated comments. Upon intervention and then every 12 hours for the next 168 hours, we also record the poster’s activity and performance (comment karma, link karma, total karma, and number of contributed posts and comments). After one week, we download the user’s complete activity history, which allows us to estimate before and after activity levels, content, interaction patterns, and karma accumulation. Similarly, we record the post’s (whether treated or treated comment’s parent) history, which gives us the complete thread content and hierarchy with timestamps and contributors’ usernames to estimate temporal patterns, interaction structure, and conversation topics. To study spillover effects, we do the same for all the post-treatment contributions by the treated user. We also download any direct private messages sent by the treated users to the two research accounts. We use an SQLite database to store the structured label and count data and a MongoDB database to store the unstructured text data. 

\section{Sample Selection with GPT-4}
\label{sec:gpt4_screen}

To screen posts and comments for sensitive topics, we use a GPT-4 screen after an initial dictionary-based filter. We use the following prompt:  “Analyze the following content and determine if it contains sensitive topics such as mental health, trauma, personal loss, serious illness, or other issues that require careful handling. Respond with `Yes' if it is sensitive and `No' if it is not sensitive. Then, provide a brief reason for your determination.” We include the text of the comment/post with the prompt but no other information, e.g. username. We note that while GPT-4 is a proprietary model, OpenAI provides robust privacy measures and encryption. Although they host the data on a cloud platform, their security protocols are advanced, reducing the risk of data breaches when handling sensitive content \cite{yao_survey_2024}. Importantly, GPT is trained on Reddit data\footnote{Source: \url{https://openai.com/index/openai- and-reddit-partnership/}} and hence, we are not sharing any content that OpenAI does not already have access to.

When implementing the field interventions, we considered using GPT-4 again to leave a public comment with each award. This would allow us to test indirect effects on observers, not just the direct effect on the recipient. However, getting GPT-4 to generate varied content-specific comments that do not reveal the use of AI but can also be believably written by humans turned out to be extremely difficult. The generated comments tended to be generic, substance-empty, and repetitive, and we were concerned that they may evoke negative reactions from the Reddit community. Hence, we left the study of indirect effects for another experiment. 

\section{Reddit Accounts for the Interventions}

\begin{figure}[ht]
\centering
\begin{subfigure}{0.48\linewidth}
\centering
\includegraphics[width=\linewidth]{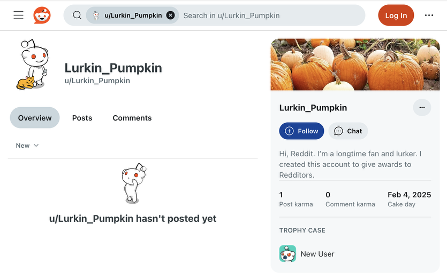}
\caption{Human account}
\end{subfigure}
\hfill
\begin{subfigure}{0.48\linewidth}
\centering
\includegraphics[width=\linewidth]{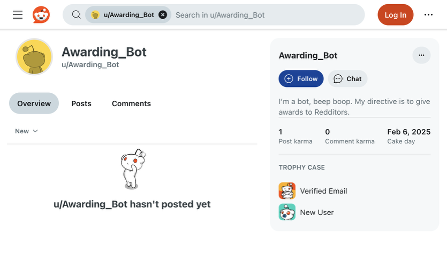}
\caption{Bot account}
\end{subfigure}
\caption{\textbf{Screenshots of the profiles of the accounts used in the experiment.}}
\label{fig:figs1}
\end{figure}

\begin{figure}[ht]
\centering
\includegraphics[width=\linewidth]{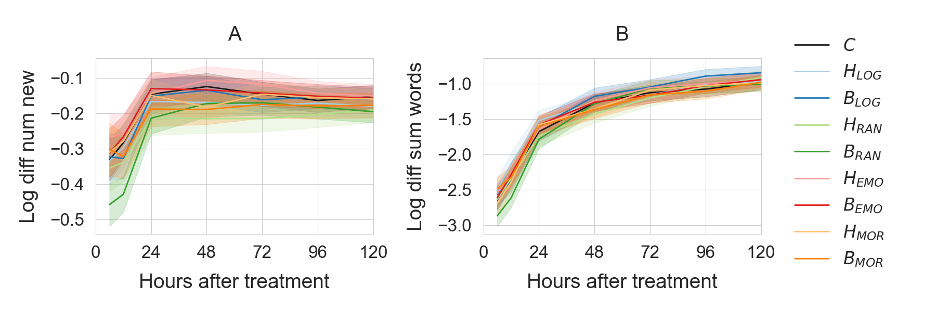}
\caption{\textbf{The direct effects of award account and rationale on the A) number and B) combined text length of new contributions the award recipient made over the 120 hours after the award.}}
\label{fig:figs2}
\end{figure}

\begin{figure}[ht]
\centering
\includegraphics[width=\linewidth]{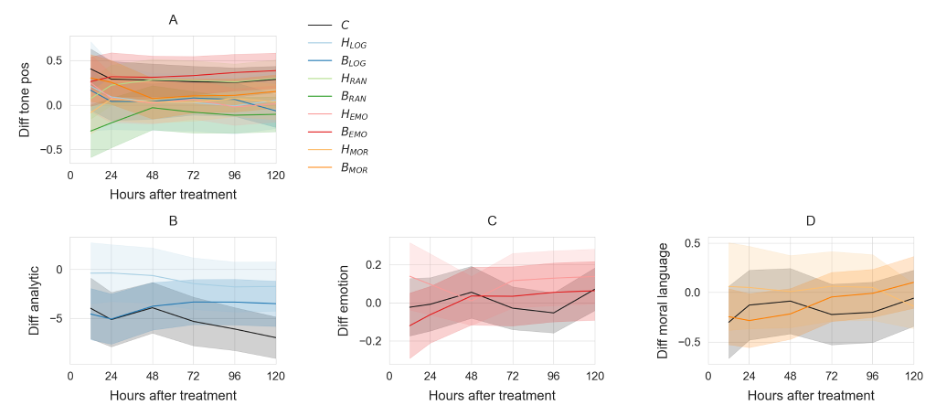}
\caption{\textbf{The direct effects of award account and rationale on the A) positive tone B) analytic language, C) emotion-related language, and D) moral language of the new contributions the award recipient made over the 120 hours after the award.}}
\label{fig:figs3}
\end{figure}

\begin{figure}[ht]
\centering
\includegraphics[width=\linewidth]{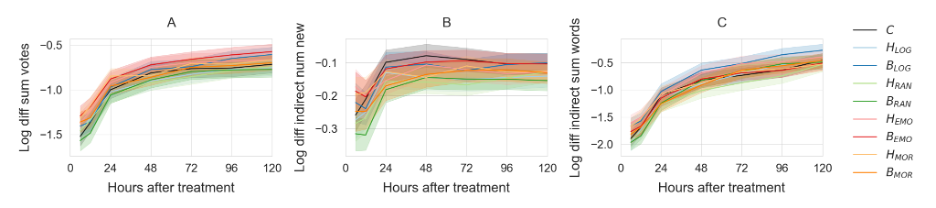}
\caption{\textbf{The indirect effects of award account and rationale on the A) total votes and the B) number and C) combined text length of the comments/replies to the new contributions the award recipient made over the 120 hours after the award.}}
\label{fig:figs4}
\end{figure}

\begin{figure}[ht]
\centering
\includegraphics[width=\linewidth]{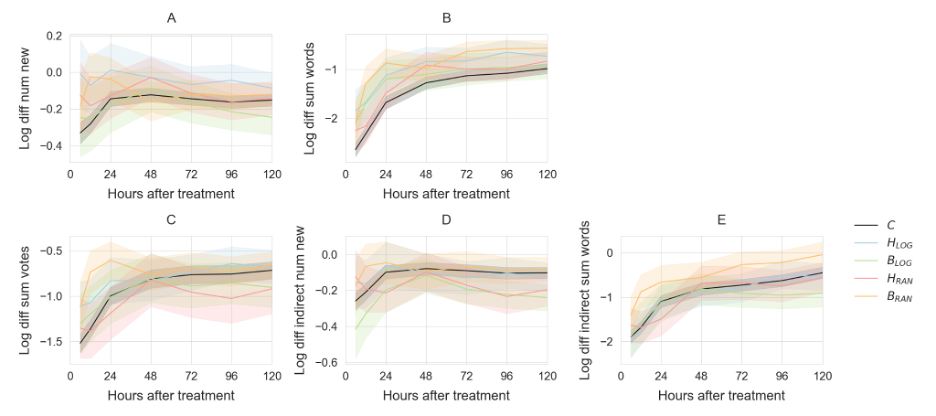}
\caption{\textbf{The direct effects of award account and rationale on the A) number and B) combined text length of new contributions the award recipient made over the 120 hours after the award for award recipients who replied to the award account with a private message, alongside the indirect effects on the C) total votes and the D) number and E) combined text length of the comments/replies these contributions received eventually.}}
\label{fig:figs5}
\end{figure}

Fig.~\ref{fig:figs1} includes screenshots of the profiles of the accounts used in the experiment. The screenshots show the username, name, avatar, background image, and profile description of the human and bot accounts, declaring their intention to give awards as a way to contribute to the community. The bot account has a profile description: “I’m a bot, beep boop. My directive is to give awards to Redditors.” The human account describes itself as “Hi, Reddit. I’m a longtime fan and lurker. I created this account to give awards to Redditors.”

\section{Ethical Considerations}

We elaborate further on three specific ethical issues: lack of informed consent, API access, and deception. The experiment relied on realistic interventions and hence, we did not obtain informed consent from the participants. We processed large volumes of text data in real time and made time-sensitive interventions. We conducted extensive tests to confirm and improve the accuracy of the dictionary- and LLM-based content filters. We also conducted a pilot trial to validate the salience of the human/bot treatment, calibrate the wording for the rationale in the messages, and estimate the expected effect strength. We used these estimates in power analyses to calculate the sample sizes for the data collection.

Reddit’s API is continuously being updated and recent changes have discontinued the functionality to automatically give awards through third-party applications.  To automate the awarding process, we employed Selenium, which allows us to programmatically control a web browser to interact with Reddit’s website as a human user would, filling in text fields and clicking buttons.

The experiment used random assignment to treatments and hence, involved a mild form of deception with respect to the “human” accounts, which used the same automated procedure as the bot accounts, and the rationale, which did not reflect the actual content of the user’s contribution. Contacting participants with a private message after the end of the experiment to debrief them is unviable because Reddit users sift through large volumes of content and are unlikely to remember the focal interaction. Instead, for debriefing, upon the publication of the study, we will make a post on the r/all subreddit that describes the experiment, lists the accounts we used for the research, and links to a preprint.

\section{Note on Results}

As it is typical for human dynamics, activity on Reddit is bursty, marked by periods of high intensity followed by periods of no engagement. We are treating contributions and thus selecting on users in an activity burst; hence, the post-treatment period is more likely to fall in a lull. We thus generally observe that after-treatment activity is lower than before-treatment activity, including for the control group. This accounts for the negative log-ratios in the figures.

\clearpage

\begin{table}[ht]
\centering
\caption{The 12 subreddits selected for the field experiments.}
\label{tab:subreddits}
\begin{tabular}{p{3.5cm}p{9cm}}
\toprule
\textbf{Subreddit} & \textbf{Focus of the community} \\
\midrule
\multicolumn{2}{l}{\textit{Personal and general advice}} \\
\midrule
r/lifeprotips       & Practical tips aimed at improving various aspects of daily life, from personal productivity to household hacks \\
r/lifehacks         & Unconventional tips and tricks to solve everyday problems efficiently, often using common items in innovative ways \\
r/personalfinance   & Advice and discussions on managing personal finances, including budgeting, saving, investing, and debt management \\
r/frugal            & Frugality and sharing strategies for saving money, budgeting, and making cost-effective decisions \\
r/careerguidance    & Advice on career development, job searching, and professional growth, helping individuals navigate their career paths \\
r/travelhacks       & Tips and tricks to make traveling more affordable, efficient, and enjoyable, covering topics like packing, booking, and navigating new places \\
\midrule
\multicolumn{2}{l}{\textit{Knowledge and information}} \\
\midrule
r/explainlikeimfive & Complex topics are broken down into simple, easy-to-understand explanations, as if explaining to a five-year-old \\
r/todayilearned     & Users share interesting and lesser-known facts they've recently discovered, starting posts with ``TIL'' (Today I Learned) \\
r/science           & Scientific discussions, news, and research across various disciplines, hosting ``Ask Me Anything'' sessions with experts to engage the public in science topics \\
r/books             & Book enthusiasts discuss literature, share recommendations, and participate in book-related conversations, including monthly book clubs and author AMAs \\
r/upliftingnews     & Positive and inspiring news stories from around the world, focusing on good deeds and successes \\
r/youshouldknow     & Important or useful information that many people might not be aware of, aiming to inform and educate \\
\bottomrule
\label{sec:table_s1}
\end{tabular}
\end{table}

\end{document}